\RequirePackage{fix-cm}
\documentclass{svjour3} 
\usepackage{amssymb,amsfonts,bm}
\usepackage{graphics,graphicx}
\usepackage{colordvi,amsmath,epsfig,color}
\usepackage[activeacute,english]{babel}
\newcommand{\sig}{\boldsymbol{\hat{\sigma}}}
\providecommand{\abs}[1]{\lvert#1\rvert}

\begin{document}

\title{The Enskog equation for confined elastic hard spheres}
\author{P. Maynar \and M. I. Garc\'ia de Soria \and J. Javier Brey}
\institute{P. Maynar \at F\'{\i}sica Te\'{o}rica, Universidad de 
Sevilla, Apartado de Correos 1065, E-41080, Sevilla, Spain \\
             \email{maynar@us.es} 
\and M. I. Garc\'ia de Soria \at F\'{\i}sica Te\'{o}rica, 
Universidad de 
Sevilla, Apartado de Correos 1065, E-41080, Sevilla, Spain \\
             \email{gsoria@us.es} 
\and J. Javier Brey \at F\'{\i}sica Te\'{o}rica, Universidad de 
Sevilla, Apartado de Correos 1065, E-41080, Sevilla, Spain \\
             \email{brey@us.es} 
}


\date{Received: date / Accepted: date}

\maketitle

\begin{abstract}
A kinetic equation for a system of elastic hard spheres or disks confined by a 
hard wall of arbitrary shape is derived. It is a generalization of the modified 
Enskog equation in which the effects of the confinement are taken into account 
and it is supposed to be valid up to moderate densities. From the equation, 
balance equations for the hydrodynamic fields are derived, identifying the 
collisional transfer contributions to the pressure tensor and heat flux. A 
Lyapunov 
functional, $\mathcal{H}[f]$, is identified. For any solution of the kinetic 
equation, $\mathcal{H}$ decays monotonically in time until the system reaches 
the inhomogeneous equilibrium distribution, that is a Maxwellian distribution 
with a the density field consistent with equilibrium statistical mechanics. 
\end{abstract}

\section{Introduction}

In 1922, Enskog introduced an equation that extends the Boltzmann equation for 
hard spheres to moderate densities \cite{e22}. By intuitive arguments, he 
modified the \emph{molecular chaos} assumption of Boltzmann, constructing an 
equation that, since then, is the paradigm of kinetic equation for moderate 
densities. It is not surprising, since it is known to be quite successful in 
describing the dynamics of dense fluids \cite{chc35,resibois}. The idea is the 
following: the Boltzmann \emph{molecular chaos} assumption can be formulated 
mathematically approximating the two-particle distribution function, $f_2$, for 
two spheres at contact and for precollisional velocities by 
\begin{equation}
f_2(\mathbf{r}+\boldsymbol{\sigma},\mathbf{v}_1,\mathbf{r},\mathbf{v}_2,t)
\approx f(\mathbf{r},\mathbf{v}_1,t)
f(\mathbf{r},\mathbf{v}_2,t), 
\end{equation}
where $\boldsymbol{\sigma}$ is a vector that joints the centers of the two 
particles at 
contact and $f$ the one-particle distribution function. Therefore, it is 
assumed that there are neither velocity correlations nor position 
correlations between the 
particles that are going 
to collide and also that $f$ does not vary appreciably in 
distance of the order of the diameter of a particle, $\sigma$. Enskog 
modified this assumption and considered 
 \begin{equation}
f_2(\mathbf{r}+\boldsymbol{\sigma},\mathbf{v}_1,\mathbf{r},\mathbf{v}_2,t)
\approx g_2(\mathbf{r}+\boldsymbol{\sigma},\mathbf{r})
f(\mathbf{r}+\boldsymbol{\sigma},\mathbf{v}_1,t)f(\mathbf{r},\mathbf{v}_2,t), 
\end{equation}
so that, still, there are not velocity correlations between the particles 
that are going to collide, but the collision does not take place with the two 
particles at the same given points. Moreover, it is assumed that the 
probability $f_2$ is modified 
with respect to the Boltzmann case by a factor $g_2$ that takes into account 
correlations between the positions of the two colliding spheres. Enskog took 
for $g_2$ the equilibrium pair correlation function at contact of a 
homogeneous fluid, calculated with the local density at the middle of the 
two particles. Under this assumption, the Enskog equation (EE) is obtained. 

Around 1970, it was shown 
that the generalization to mixtures of the EE was not consistent 
with the laws of irreversible thermodynamics, Onsager's reciprocal relations 
were violated \cite{gbp71}. Later on, it was realized that the problem could 
be solved by means of a 
modification of the standard EE, consisting in taking $g_2$ to be the pair 
correlation 
function at contact of an inhomogeneous fluid at equilibrium in the presence 
of a force field such that the equilibrium density of this reference 
system be the instantaneous actual density 
field, $n(\mathbf{r},t)$. This new equation is called the modified Enskog 
equation (MEE) \cite{vbe73}. Intuitively, this new hypothesis seems appealing, 
since it takes into account the spatial correlations between two particles 
in a non-uniform local equilibrium state, while in the standard framework 
non-uniformities are only taken into account to a certain extend. Moreover, 
the MEE has several advantages against the standard EE: a) It can be 
derived from the Liouville equation assuming that, for all times, the 
N-particle distribution function, $\rho_N$, is such that there are not velocity 
correlations, although all hard spheres overlap exclusions are taken into 
account \cite{lps67}. b) An $\mathcal{H}$-theorem can be derived for the MEE 
\cite{rPRL78,r78}. c) In the presence of an external field, the MEE yields the 
correct single-particle equilibrium distribution function, whereas the standard 
EE does not \cite{vb83}. 

Until now, as far as we know, the EE (in its two versions) has been considered 
for an 
infinite system or with periodic boundary conditions \cite{chc35,resibois}. 
In particular, the derivation of the 
$\mathcal{H}$-theorem \cite{rPRL78,r78} is restricted to periodic boundary 
conditions. A priori, it seems difficult to deal with the 
excluded volume effects caused by both, the boundary and the particles. Very 
recently, a kinetic equation for a dilute system composed of 
hard spheres that takes into account the effect of confinement was 
proposed \cite{bmg16}. The particles are confined between two parallel 
plates separated a distance smaller than two particle diameters. Here, by 
extending these ideas, the MEE is 
formulated taking into account the effects of arbitrary confinement (a hard 
wall with arbitrary shape). In fact, as it will be seen along the paper, it 
has the same conceptual advantages that the MEE: it can be derived from the 
Liouville equation under some approximations, it is consistent with the 
equilibrium distribution 
function, and an $\mathcal{H}$-theorem can be derived. Moreover, balance 
equations for the hydrodynamic fields will be obtained and the main differences 
with the ones from the MEE will be discussed. 

In the last decades, the study of confined fluids has attracted a great deal of 
attention, mostly focused on equilibrium and phase transition properties 
\cite{tgr92,dh95,sl96,sl97,fls12,tme87,rslt}. On the other hand, few 
non-equilibrium results for confined systems seem to be well 
established in the context of a general theory. In this sense, the equation 
formulated in this paper goes in the direction of filling this gap, as 
it let us study the dynamic of dense confined systems in the hard 
spheres case. It also opens the possibility of studying new 
questions as, for example, existence of hydrodynamics or, if this is the case, 
how the hydrodynamic equations are modified. These effects are expected to be 
particularly 
important in situations of strong confinement, i.e. when 
the size of the particles is of the order of the geometrical parameters 
describing the confinement, 
as in the previous example of the two parallel plates. 

The paper is organized as follows: in section \ref{section2} the MEE for a 
general confinement is derived. It is shown that the equation admits 
the Maxwellian equilibrium distribution with the density profile predicted by 
statistical mechanics. It is also shown that the equation reduces to the one 
introduced in \cite{bmg16} in the appropriated limit. In section 
\ref{section3}, 
balance equations for the hydrodynamic fields are deduced, while in section 
\ref{section4} the $\mathcal{H}$-theorem is proved. Finally, in section 
\ref{section5} some concluding remarks are formulated.

\section{Kinetic equation}\label{section2}

The model we consider is an ensemble of $N$ elastic hard spheres 
($d=3$) or disks 
($d=2$), of mass $m$ and diameter $\sigma$. 
The particles are confined inside a volume $V$ with a boundary $\partial V$. 
In principle, the shape of the boundary 
surface is arbitrary and can have corners, but it is assumed that it is such 
that particles can explore all the volume. Let us mention that, when we refer 
to $V$, we mean the accessible volume to the centers of the particles (the 
distance between 
any point of $\partial V$ and the actual wall is, hence, $\sigma/2$). At a 
given time $t$, the state of 
the system is given by the positions and velocities of the $N$ particles, 
$\{\mathbf{R}_1(t),\mathbf{V}_1(t), \dots,\mathbf{R}_N(t),\mathbf{V}_N(t)\}$, 
where $\mathbf{R}_i$ and $\mathbf{V}_i$ are the position and velocity of 
particle $i$ respectively. 
The dynamics consists of free streaming until there is an encounter between 
two particles, or of a particle with the wall. Suppose that there is a 
collision 
of two particles, say particle $1$ and $2$, with velocities 
$\mathbf{V}_1$ and $\mathbf{V}_2$ respectively, the postcollisional velocities 
are 
\begin{eqnarray}
\mathbf{V}_1'\equiv b_{\sig}\mathbf{V}_1=\mathbf{V}_1-
(\mathbf{V}_{12}\cdot\sig)\sig, \\
\mathbf{V}_2'\equiv b_{\sig}\mathbf{V}_2=\mathbf{V}_2+
(\mathbf{V}_{12}\cdot\sig)\sig, 
\end{eqnarray}
where $\mathbf{V}_{12}\equiv\mathbf{V}_1-\mathbf{V}_2$ is the relative velocity, 
and $\sig$ a unitary vector joining the centers of the two particles at 
contact (from $2$ 
to $1$). We have also introduced the operator $b_{\sig}$ that changes 
functions of $\mathbf{V}_1, \mathbf{V}_2$ to the same functions of the 
scattered velocities, i.e. 
$b_{\sig} g(\mathbf{V}_1,\mathbf{V}_2)\equiv g(\mathbf{V}_1',\mathbf{V}_2')$ for 
any arbitrary function $g$. When a 
particle collides with the wall at $\mathbf{r}\in\partial V$ with velocity 
$\mathbf{V}$, it 
experiments an elastic reflexion, and its velocity after the collision is 
\begin{equation}\label{opbe}
b_e(\mathbf{r})\mathbf{V}=\mathbf{V}-2[\mathbf{V}\cdot\mathbf{N}(\mathbf{r})]
\mathbf{N}(\mathbf{r}).  
\end{equation}
Here we have introduced the operator $b_e(\mathbf{r})$ and the unitary vector 
normal to the surface at $\mathbf{r}$ with an outward orientation, 
$\mathbf{N}(\mathbf{r})$. As the surface can have corners, the vectorial field, 
$\mathbf{N}(\mathbf{r}): \mathbf{r}\in\partial V \to \mathbb{R}^d$, may have a 
finite number of discontinuities. Let us mention that, although the model can 
be easily generalized to other collision rules and other kind of ``hard'' 
interactions with the wall, we will restrict ourselves to this simple case, 
because the kinetic equation that will be derived includes all the new 
ingredients that we want to analyze. 

\begin{figure}
\begin{center}
\includegraphics[angle=0,width=0.5\linewidth,clip]{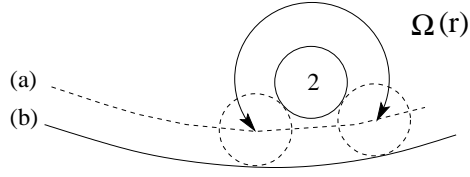}
\end{center}
\caption{Schematic representation of the restricted solid angle for particle 2 
locate at $\mathbf{r}$, 
$\Omega(\mathbf{r})$. Line (a) is the boundary of $V$, while line (b) is the 
actual wall. }\label{fo}
\end{figure}

Now, the objective is to derive a kinetic equation for this model, i.e. a 
closed equation for the one-particle distribution function, 
$f(\mathbf{r},\mathbf{v},t)$. This function is defined as usual in kinetic 
theory, so that 
$\int_{V_1}d\mathbf{r}\int_{W_1}d\mathbf{v}f(\mathbf{r},\mathbf{v},t)$ is 
the mean number of particles with positions inside the volume $V_1$  and 
velocities inside $W_1$ at time $t$, for any of such volumes. 
In \cite{Cercignani}, the BBGKY hierarchy is 
derived for the present model, taking into account the hard wall. The first 
equation of the hierarchy relates the one-particle distribution function with 
the two-particle distribution function, 
$f_2(\mathbf{r}_1,\mathbf{v}_1,\mathbf{r}_2,\mathbf{v}_2,t)$, which is defined 
in such a way that $\int_{V_1}d\mathbf{r}_1\int_{W_1}d\mathbf{v}_1
\int_{V_2}d\mathbf{r}_2\int_{W_2}d\mathbf{v}_2
f_2(\mathbf{r}_1,\mathbf{v}_1,\mathbf{r}_2,\mathbf{v}_2,t)$ is the mean number 
of pairs of particles such that particle 1 is in $V_1$ with velocity in 
$W_1$, while particle 2 is in $V_2$ with velocity in $W_2$ at 
time $t$. The equation is

\begin{equation}\label{bbgky1}
\left(\frac{\partial}{\partial t}
+\mathbf{v}_2\cdot\frac{\partial}{\partial\mathbf{r}}\right)
f(\mathbf{r},\mathbf{v}_2,t)=J[f_2], 
\end{equation}

with
\begin{equation}\label{C1}
J[f_2]=\sigma^{d-1}\int d\mathbf{v}_1\int d\sig
\abs{\mathbf{v}_{12}\cdot\sig}[\theta(\mathbf{v}_{12}\cdot\sig)b_{\sig}-
\theta(-\mathbf{v}_{12}\cdot\sig)]
f_2(\mathbf{r}+\boldsymbol{\sigma},\mathbf{v}_1,\mathbf{r},\mathbf{v}_2,t). 
\end{equation}
Here $d\sig$ is the solid angle element for 
$\sig$, and $\theta$ is the Heaviside step function. The integration is over 
the complete velocity space and the total solid angle for dimension $d$, 
$\Omega_d$. Looking further at $J[f_2]$, it is seen that, closed to 
the boundary, it can happen that $\mathbf{r}\in V$ while 
$\mathbf{r}+\boldsymbol{\sigma}\notin V$ for certain $\sig\in\Omega_d$. Of 
course, 
$f_2(\mathbf{r}+\boldsymbol{\sigma},\mathbf{v}_1,\mathbf{r},\mathbf{v}_2,t)=0$ 
if $\mathbf{r}+\boldsymbol{\sigma}\notin V$. This is due to the fact that, if 
the initial 
condition is such that all the particles are inside the volume, the dynamic 
conserves this property. 
Let us remark that Eq. (\ref{bbgky1}) has no terms corresponding to 
particle-wall collisions because they are included in the boundary conditions 
of the distribution functions $f$ and $f_2$. In fact, it is possible to 
formulate an equivalent equation with a new term that incorporates the 
collisions with the walls. Then, the equation can be split into a regular part 
and a singular part. The regular part is Eq. (\ref{bbgky1}) and the 
singular part are the boundary conditions. This has been explicitely done in 
Ref. \cite{bgm17} for the special geometry of two paralell walls separated a 
distance smaller than twice the diameter of the particles. 
In the following, it will be convenient 
to express $J[f_2]$ in terms of the configurations that are actually 
allowed. This can be done by taking into account that, for fixed $\mathbf{r}$, 
only a restricted solid angle, $\Omega(\mathbf{r})$, is possible, in such a 
way that $\mathbf{r}+\boldsymbol{\sigma}\in V$ if and only if 
$\sig\in\Omega(\mathbf{r})$ (see Fig. \ref{fo}). Then, $J[f_2]$ can be 
expressed as

\begin{equation}\label{C2}
J[f_2]=\sigma^{d-1}\int d\mathbf{v}_1\int_{\Omega(\mathbf{r})} d\sig
\abs{\mathbf{v}_{12}\cdot\sig}[\theta(\mathbf{v}_{12}\cdot\sig)b_{\sig}-
\theta(-\mathbf{v}_{12}\cdot\sig)]
f_2(\mathbf{r}+\boldsymbol{\sigma},\mathbf{v}_1,\mathbf{r},\mathbf{v}_2,t). 
\end{equation}

\vspace*{0.5cm}
\vspace*{0.5cm}

Until now, Eq. (\ref{bbgky1}) is not closed. Nevertheless, let us see that, in 
a similar fashion that in the non-confined case \cite{vbe73,lps67}, $f_2$ can 
be expressed as a functional of $f$ under some mathematically well defined 
approximations. Let us \emph{assume} that, for the evaluation of 
some reduced distribution functions and for precollisional configurations, 
the $N$-particle probability distribution function, $\rho_N$, can be 
approximated in the form 

\begin{equation}\label{rhoN}
\rho_N(\Gamma,t)\approx\frac{\Theta(\mathbf{r}_1,\dots,\mathbf{r}_N)}{\phi(t)}
\prod_{n=1}^NW(\mathbf{r}_n,\mathbf{v}_n,t), 
\end{equation}
for all times and for $\mathbf{r}_n\in V, n=1,2,\dots,N$, where 
$\Gamma\equiv(\mathbf{r}_1,\mathbf{v}_1,\dots,\mathbf{r}_N,\mathbf{v}_N)$, and 
\begin{equation}
\Theta(\mathbf{r}_1,\dots,\mathbf{r}_N)\equiv\prod_{i=1}^N\prod_{j>i}
\theta(\abs{\mathbf{r}_i-\mathbf{r}_j}-\sigma), 
\end{equation}
that vanishes for overlapped configurations. 
The function $W$ can be considered to be normalized, i.e. 
$\int d\mathbf{r}\int d\mathbf{v}W(\mathbf{r},\mathbf{v},t)=1$, and the 
function of time, 
\begin{equation}\label{phi}
\phi(t)=\int d\mathbf{r}_1w(\mathbf{r}_1,t)\dots\int d\mathbf{r}_N
w(\mathbf{r}_N,t)\Theta(\mathbf{r}_1,\dots,\mathbf{r}_N), 
\end{equation}
with 
\begin{equation}
w(\mathbf{r},t)\equiv\int d\mathbf{v}W(\mathbf{r},\mathbf{v},t), 
\end{equation}
arises as a normalization factor of $\rho_N$. The integrals in the space 
variable are supposed to be over the confining volume, $V$. 
The crucial assumption in Eq. (\ref{rhoN}) is that the 
velocity dependence of $\rho_N$ enters only through the function 
$W(\mathbf{r},\mathbf{v},t)$ in a factorized way, keeping the exact property 
that $\rho_N$ must vanish for overlapped configurations due to 
$\Theta(\mathbf{r}_1,\dots,\mathbf{r}_N)$. Although the form given by Eq. 
(\ref{phi}) is exact at 
equilibrium (in the canonical ensemble $W$ is the Maxwellian distribution), it 
can only be 
an approximation for out of equilibrium systems \cite{lps67,rPRL78,r78}. 
Concretely, it is 
a good approximation for precollisional configurations but, as it will be 
shown, then it can not be valid for postcollisional configurations. 

Now, let us use the approximation given by Eq. (\ref{rhoN}) to evaluate the 
one-particle distribution function, $f(\mathbf{r},\mathbf{v},t)$, and the 
two-particle distribution function at contact for precollisional velocities, 
$f_2(\mathbf{r}+\boldsymbol{\sigma},\mathbf{v}_1,\mathbf{r},\mathbf{v}_2,t)$ 
with $\sig\cdot\mathbf{v}_{12}<0$. This is, in fact, the part 
of the distribution that we need as it is the part that appears in Eq. 
(\ref{C2}). The one-particle distribution function is 
\begin{equation}\label{fW}
f(\mathbf{r}_1,\mathbf{v}_1,t)=\frac{N}{\phi(t)}W(\mathbf{r}_1,\mathbf{v}_1,t)
\int d\mathbf{r}_2w(\mathbf{r}_2,t)\dots\int d\mathbf{r}_Nw(\mathbf{r}_N,t)
\Theta(\mathbf{r}_1,\dots,\mathbf{r}_N), 
\end{equation}
and the density field, 
\begin{equation}\label{dw}
n(\mathbf{r}_1,t)=\frac{N}{\phi(t)}w(\mathbf{r}_1,t)
\int d\mathbf{r}_2w(\mathbf{r}_2,t)\dots\int d\mathbf{r}_Nw(\mathbf{r}_N,t)
\Theta(\mathbf{r}_1,\dots,\mathbf{r}_N). 
\end{equation}
Here it is seen that the density, $n$, is a functional of $w$ (note that $\phi$ 
is also a functional of $w$ by Eq. (\ref{phi})). 
In fact, according to a theorem of density functional theory that establishes 
that, for a fluid in equilibrium in the presence of an external potential, 
$E_p(\mathbf{r})$, there is a one to one correspondence between the external 
potential and 
the density field \cite{e79}, it can be expected that $w$ is also a 
functional of $n$, i.e. 
\begin{equation}\label{wF}
w(\mathbf{r},t)=\mathcal{Y}(\mathbf{r},t|n). 
\end{equation}
This is because the functional given by Eq. (\ref{dw}) is the same 
that the one that appears in the context of density functional theory by 
making the substitution 
$w(\mathbf{r})\leftrightarrow e^{-\frac{E_p(\mathbf{r})}{T}}$, where $T$ is the 
temperature, that is an arbitrary parameter, and the Boltzmann constant, $k_B$, 
has been taken to be unity. In any case, Eq. (\ref{wF}) can be taken 
as an additional assumption. 
The two-particle distribution function at contact 
for precollisional velocities can be expressed in the form 
\begin{equation}\label{f2}
f_2(\mathbf{r}+\boldsymbol{\sigma},\mathbf{v}_1,\mathbf{r},\mathbf{v}_2,t)=
g_2(\mathbf{r}+\boldsymbol{\sigma},\mathbf{r}\lvert n)
f(\mathbf{r}+\boldsymbol{\sigma},\mathbf{v}_1,t)f(\mathbf{r},\mathbf{v}_2,t), 
\end{equation}
valid for $\sig\cdot\mathbf{v}_{12}<0$, 
where $g_2$ is the pair correlation function at contact, defined as 
\begin{equation}
g_2(\mathbf{r}+\boldsymbol{\sigma},\mathbf{r}\lvert n)=
\frac{n_2(\mathbf{r}+\boldsymbol{\sigma},\mathbf{r},t)}
{n(\mathbf{r}+\boldsymbol{\sigma},t)n(\mathbf{r},t)}, 
\end{equation}
with 
\begin{eqnarray}\label{n2w}
n_2(\mathbf{r}+\boldsymbol{\sigma},\mathbf{r},t)
=\frac{N(N-1)}{\phi(t)} w(\mathbf{r}+\boldsymbol{\sigma},t)w(\mathbf{r},t)
\int d\mathbf{r}_3w(\mathbf{r}_3,t)\dots
\nonumber\\
\dots\int d\mathbf{r}_Nw(\mathbf{r}_N,t)
\Theta(\mathbf{r}+\boldsymbol{\sigma},\mathbf{r},\mathbf{r}_3,
\dots,\mathbf{r}_N).
\end{eqnarray}
As $w$ is a functional of the density, $n_2$ and consequently $g_2$ are also 
functionals of the density (in the notation 
introduced for $g_2$, this is explicitly indicated). Moreover, the functional 
$g_2$ is the same as the one associated to a 
system in equilibrium at temperature $T$, in the presence of an external force, 
$\mathbf{F}=-\frac{\partial E_p(\mathbf{r})}{\partial\mathbf{r}}$, such that 
$w(\mathbf{r})\propto e^{-\frac{E_p(\mathbf{r})}{T}}$ \cite{vbe79}. Let us also 
remark that the two-particle distribution function at contact for 
postcollisional velocities can be consistently calculated in the framework of 
approximation given by Eq. (\ref{f2}). Due to the conservation of probability 
in a collision, it can be written in the form \cite{l96}
\begin{equation}
f_2(\mathbf{r}+\boldsymbol{\sigma},\mathbf{v}_1,\mathbf{r},\mathbf{v}_2,t)=
g_2(\mathbf{r}+\boldsymbol{\sigma},\mathbf{r}\lvert n)b_{\sig}
f(\mathbf{r}+\boldsymbol{\sigma},\mathbf{v}_1,t)f(\mathbf{r},\mathbf{v}_2,t), 
\end{equation}
for $\sig\cdot\mathbf{v}_{12}>0$. Here it is seen that Eq. (\ref{rhoN}) 
is clearly inconsistent for postcollisional configurations. 

By substituting the factorized form of the two-particle distribution function, 
Eq. (\ref{f2}), into Eq. (\ref{C2}), the collisional contribution is expressed 
in terms of $f$. The obtained evolution equation for $f$ is then

\begin{equation}\label{enskog}
\left(\frac{\partial}{\partial t}
+\mathbf{v}_2\cdot\frac{\partial}{\partial\mathbf{r}}\right)
f(\mathbf{r},\mathbf{v}_2,t)=J_E[f\lvert f], 
\end{equation}
with
\begin{eqnarray}\label{Jenskog}
J_E[f\lvert f]=\sigma^{d-1}\int d\mathbf{v}_1\int_{\Omega(\mathbf{r})} 
d\sig\abs{\mathbf{v}_{12}\cdot\sig}[\theta(\mathbf{v}_{12}\cdot\sig)b_{\sig}-
\theta(-\mathbf{v}_{12}\cdot\sig)]
\nonumber\\
g_2(\mathbf{r}+\boldsymbol{\sigma},\mathbf{r}\lvert n)
f(\mathbf{r}+\boldsymbol{\sigma},\mathbf{v}_1,t)f(\mathbf{r},\mathbf{v}_2,t), 
\end{eqnarray}
to be solved with the boundary conditions 
\begin{equation}\label{bc}
f(\mathbf{r},\mathbf{v},t)=f(\mathbf{r},b_e\mathbf{v},t), 
\quad\forall\mathbf{r}\in\partial V, \quad\forall t,\quad\forall\mathbf{v} 
\quad\textrm{with $\mathbf{v}\cdot\mathbf{N}(\mathbf{r})>0$}. 
\end{equation}

Eq. (\ref{enskog}) is the closed equation for the one-particle distribution 
function we were looking for. It describes the dynamics of a system of hard 
spheres or disks confined by hard walls of arbitrary shape, and can be 
considered as the starting point to tackle other questions, such as the 
derivation of hydrodynamics, modification of transport coefficients, etc... 
As the MEE, it is expected to be valid for moderate 
densities. The difference between Eq. (\ref{enskog}) and the MEE 
resides in the fact that the region of integration of the solid angle depends 
on $\mathbf{r}$. In the bulk, the two equations coincide but, closed to the 
boundary, the possible solid angles in Eq. (\ref{enskog}) are restricted by the 
fact that particles must be inside the volume $V$, i.e. $\mathbf{r}\in V$ and 
$\mathbf{r}+\boldsymbol{\sigma}\in V$. Of course, the functional $g_2$ depends 
also on the shape of the container. 


By direct substitution, it is shown that Eq. (\ref{enskog}) admits a stationary 
or equilibrium solution of the form
\begin{equation}\label{fe}
f_e(\mathbf{r},\mathbf{v})=n_e(\mathbf{r})\chi_M(v,T), 
\end{equation}
where $\chi_M(v,T)$ is a Maxwellian distribution of temperature $T$
\begin{equation}\label{max}
\chi_M(v,T)=\frac{e^{-\frac{v^2}{v_0^2}}}{\pi^{d/2}v_0^d}, 
\end{equation}
with $v_0$ being the thermal velocity defined through 
$T\equiv\frac{m}{2}v_0^2$. The temperature is defined as usual, 
$\frac{d}{2}nT\equiv\int d\mathbf{v}\frac{m}{2}v^2f$. In effect, when Eq. 
(\ref{fe}) is substituted into Eq. (\ref{enskog}), the velocity dependence is 
eliminated and the equilibrium density must fulfill
\begin{equation}\label{eqne}
\frac{\partial}{\partial\mathbf{r}}\ln n_e(\mathbf{r})=-\sigma^{d-1}
\int_{\Omega(\mathbf{r})}g_2(\mathbf{r}+\boldsymbol{\sigma},\mathbf{r}\lvert n_e)
n_e(\mathbf{r}+\boldsymbol{\sigma})\sig, 
\end{equation}
that is the first equation of the BGY hierarchy \cite{HansenMcDonald}. Hence, 
the distribution function, $f_e$, is consistent with the known properties of 
equilibrium statistical mechanics. Note also that, as $\chi_M$ depends on 
$\abs{\mathbf{v}}$, the boundary conditions given by Eq. (\ref{bc}) are 
automatically satisfied. 

Finally, let us remark that, for low densities, Eq. (\ref{enskog}) reduces to a 
much simpler form. In effect, the first term in the Mayer expansion of 
the pair correlation function is 
$g_2(\mathbf{r}_1,\mathbf{r}_2\lvert n)\sim
\theta(\abs{\mathbf{r}_1-\mathbf{r}_2}-\sigma)$, so that, in 
this limit, Eq.(\ref{enskog}) takes the form
 \begin{equation}\label{be}
\left(\frac{\partial}{\partial t}
+\mathbf{v}_2\cdot\frac{\partial}{\partial\mathbf{r}}\right)
f(\mathbf{r},\mathbf{v}_2,t)=J_{BE}[f\lvert f], 
\end{equation}
with
\begin{eqnarray}\label{Jbe}
J_{BE}[f\lvert f]=\sigma^{d-1}\int d\mathbf{v}_1\int_{\Omega(\mathbf{r})} 
d\sig\abs{\mathbf{v}_{12}\cdot\sig}[\theta(\mathbf{v}_{12}\cdot\sig)b_{\sig}-
\theta(-\mathbf{v}_{12}\cdot\sig)]
\nonumber\\
f(\mathbf{r}+\boldsymbol{\sigma},\mathbf{v}_1,t)f(\mathbf{r},\mathbf{v}_2,t).  
\end{eqnarray}
The collision operator has some similarities with the Boltzmann collision 
operator, because it does not contain the correlation function, $g_2$, but the 
dependence of $f$ on distances of order $\sigma$ is still important, as in 
Enskog. Considering a confinement between two parallel walls separated a 
distance smaller than two particle diameters, and assuming that the 
distribution does not vary on distance of order $\sigma$ in the directions 
parallels to the planes, the equation analyzed in reference \cite{bmg16} is 
obtained. Let us remark that the equation of reference \cite{bmg16} describes 
correctly the equilibrium 
properties for densities beyond Boltzmann and also some studied 
nonequilibrium 
dynamical properties \cite{bgm17}. Hence, it seems that, when the density is 
not so high, Eq. (\ref{be}) represents 
a good starting point for the study of confined fluids, in a more simplified 
way that with Eq. (\ref{enskog}). Note that the Boltzmann equation is obtained 
in the Grad limit, where it is fair to approximate 
$f(\mathbf{r}+\boldsymbol{\sigma},\mathbf{v},t)\sim f(\mathbf{r},\mathbf{v},t)$ 
and $\Omega(\mathbf{r})=\Omega_d$. Clearly, this limit has no sense when the 
geometrical constraints due to the boundary are in some direction of the order 
of the size of the particles. 

\section{Balance equation}\label{section3}

The hydrodynamic fields are defined as usual in kinetic theory, as the first 
velocity moments of the one-particle distribution function
\begin{eqnarray}
n(\mathbf{r},t)&=&\int d\mathbf{v}f(\mathbf{r},\mathbf{v},t),\\
n(\mathbf{r},t)\mathbf{u}(\mathbf{r},t)
&=&\int d\mathbf{v}\mathbf{v}f(\mathbf{r},\mathbf{v},t), \\
\frac{d}{2}n(\mathbf{r},t)T(\mathbf{r},t)
&=&\frac{m}{2}\int d\mathbf{v}[\mathbf{v}-\mathbf{u}(\mathbf{r},t)]^2
f(\mathbf{r},\mathbf{v},t), 
\end{eqnarray}


By taking velocity moments in the first equation of the BBGKY, Eq. 
(\ref{bbgky1}), formal relations between the hydrodynamic fields and the fluxes 
are obtained 
\begin{eqnarray}\label{eh1}
\frac{\partial}{\partial t}n+\frac{\partial}{\partial\mathbf{r}}
\cdot(n\mathbf{u})=0, \\
\label{eh2}
\frac{\partial}{\partial t}(nu_i)+\frac{\partial}{\partial\mathbf{r}}
\cdot(nu_i\mathbf{u})+\frac{1}{m}\frac{\partial}{\partial\mathbf{r}}\cdot 
P^{(k)}=\int d\mathbf{v}v_iJ[f_2],\\
\label{eh3}
\frac{\partial}{\partial t}\left(\frac{d}{2}nT+\frac{m}{2}nu^2\right)+
\frac{\partial}{\partial\mathbf{r}}\cdot
\left(\frac{d}{2}nT\mathbf{u}+\frac{m}{2}nu^2\mathbf{u}\right)\nonumber\\
+\frac{\partial}{\partial\mathbf{r}}\cdot
\left(\mathbf{u}\cdot P^{(k)}+\mathbf{q}^{(k)}\right)
=\frac{m}{2}\int d\mathbf{v}v^2J[f_2], 
\end{eqnarray}
where we have introduced the kinetic pressure tensor
\begin{equation}\label{pk}
P_{ij}^{(k)}(\mathbf{r},t)=m\int d\mathbf{v}[v_i-u_i(\mathbf{r},t)]
[v_j-u_j(\mathbf{r},t)]f(\mathbf{r},\mathbf{v},t), 
\end{equation}
and the kinetic heat flux 
\begin{equation}\label{qk}
\mathbf{q}^{(k)}(\mathbf{r},t)=\frac{m}{2}
\int d\mathbf{v}[\mathbf{v}-\mathbf{u}(\mathbf{r},t)]^2
[\mathbf{v}-\mathbf{u}(\mathbf{r},t)]f(\mathbf{r},\mathbf{v},t). 
\end{equation}
At first sight, it seems that Eqs. (\ref{eh2}) and (\ref{eh3}) are not 
associated to conserved quantities, due to the collisional terms, i.e. the 
terms that involve $J$. But, in fact, 
this is not the case because, as it will be shown, these terms can be 
transformed into the 
divergence of a quantity that is associated to the collisional flux of momentum 
and energy. From a physical point of view the picture is the following: there 
is flux of momentum and energy through a given surface due to particles that 
cross the surface and due to collisions between particles (the two particles 
are in opposite sites of the surface, do not cross the surface, but interchange 
momentum and energy due to collisions). The first contribution to the fluxes 
are the kinetic fluxes defined above, while the second contribution can be 
evaluated by kinetic theory arguments, just by counting collisions and taken 
into account the corresponding contribution to the fluxes. This is done in 
Appendix \ref{cf}, obtaining the collisional contribution to the pressure tensor
\begin{eqnarray}\label{Pijc}
P_{ij}^{(c)}(\mathbf{r},t)
=\frac{m}{2}\sigma^d\int d\mathbf{v}_1\int d\mathbf{v}_2\int\int_{\Sigma} 
d\lambda 
d\sig\theta(-\mathbf{v}_{12}\cdot\sig)
\nonumber\\
f_2[\mathbf{r}_1(\lambda,\sig),
\mathbf{v}_1,\mathbf{r}_2(\lambda,\sig),\mathbf{v}_2,t]
(\mathbf{v}_{12}\cdot\sig)^2\hat{\sigma_i}\hat{\sigma_j}, 
\end{eqnarray}
and the collisional contribution to the heat flux 
\begin{eqnarray}\label{qc}
\mathbf{q}^{(c)}(\mathbf{r},t)=-\mathbf{u}\cdot P^{(c)}
+\frac{m}{4}\sigma^d\int d\mathbf{v}_1\int d\mathbf{v}_2\int\int_{\Sigma} 
d\lambda 
d\sig\theta(-\mathbf{v}_{12}\cdot\sig)
\nonumber\\
f_2[\mathbf{r}_1(\lambda,\sig),
\mathbf{v}_1,\mathbf{r}_2(\lambda,\sig),\mathbf{v}_2,t]
(\mathbf{v}_1+\mathbf{v}_2)\cdot\sig (\mathbf{v}_{12}\cdot\sig)^2\sig.
\end{eqnarray}
Here, we have introduced the functions 
\begin{eqnarray}
\mathbf{r}_1(\lambda,\sig)&=&\mathbf{r}+\lambda\boldsymbol{\sigma}, \\
\mathbf{r}_2(\lambda,\sig)&=&\mathbf{r}-(1-\lambda)\boldsymbol{\sigma},  
\end{eqnarray}
and the region of integration in the $(\lambda,\sig)$ space 
\begin{equation}
\Sigma=\{(\lambda,\sig)\lvert\sig\in\Omega_d\quad\&\quad 0\le\lambda\le 1
\quad\&\quad\mathbf{r}_1(\lambda,\sig)\in V
\quad\&\quad\mathbf{r}_2(\lambda,\sig)\in V \}. 
\end{equation}
Although it can be argued that Eqs.  (\ref{Pijc}) and (\ref{qc}) are proposed 
on the basis of intuitive arguments, they play the desired rule because, as it 
is shown in Appendix \ref{dcf}, they fulfill 
\begin{eqnarray}\label{jv}
\int d\mathbf{v}m\mathbf{v}J[f_2]&=&
-\frac{\partial}{\partial\mathbf{r}}\cdot P^{(c)}, \\
\label{jv2}
\frac{m}{2}\int d\mathbf{v}v^2J[f_2]&=&-\frac{\partial}{\partial\mathbf{r}}\cdot
\left(\mathbf{q}^{(c)}+\mathbf{u}\cdot P^{(c)}\right).  
\end{eqnarray}
Finally, by substituting Eqs. (\ref{jv}) and (\ref{jv2}) into Eqs. (\ref{eh2}) 
and (\ref{eh3}) respectively, it is obtained
\begin{eqnarray}\label{eh1b}
\frac{\partial}{\partial t}n+\frac{\partial}{\partial\mathbf{r}}
\cdot(n\mathbf{u})=0, \\
\label{eh2b}
\frac{\partial}{\partial t}(nu_i)+\frac{\partial}{\partial\mathbf{r}}
\cdot(nu_i\mathbf{u})+\frac{1}{m}\frac{\partial}{\partial\mathbf{r}}\cdot 
P=0,\\
\label{eh3b}
\frac{\partial}{\partial t}\left(\frac{d}{2}nT+\frac{m}{2}nu^2\right)+
\frac{\partial}{\partial\mathbf{r}}\cdot
\left(\frac{d}{2}nT\mathbf{u}+\frac{m}{2}nu^2\mathbf{u}\right)\nonumber\\
+\frac{\partial}{\partial\mathbf{r}}\cdot
\left(\mathbf{u}\cdot P+\mathbf{q}\right)
=0, 
\end{eqnarray}
where the total pressure tensor and heat flux have been introduced 
\begin{eqnarray}
P&=&P^{(k)}+P^{(c)}, \\
\mathbf{q}&=&\mathbf{q}^{(k)}+\mathbf{q}^{(c)}. 
\end{eqnarray}
The structure of Eqs. (\ref{eh1b})-(\ref{eh3b}) clearly shows that they 
are associated to conserved quantities, and they are the starting point to 
derive hydrodynamic equations. If the one and two-particle distribution 
functions are expressed in terms of the hydrodynamic fields and their 
gradients, the kinetic and collisional fluxes are expressed in the same way, 
and closed equations for the hydrodynamic fields are obtained. 

If the Enskog equation is taken as the starting point in the derivation of the 
hydrodynamic equations (instead of the first 
equation of the BBGKY hierarchy), the same equations are obtained, Eqs. 
(\ref{eh1b})-(\ref{eh3b}). The expression for the kinetic fluxes are the same, 
Eqs. (\ref{pk}) and (\ref{qk}), while the collisional contribution is slightly 
modified. Specifically, the expression for the collisional fluxes are given by 
Eqs. (\ref{Pijc}) and (\ref{qc}), but substituting the 
exact two-particle distribution, $f_2$, by the approximate 
factorized form given by Eq. (\ref{f2}). The collisional contribution of the 
fluxes coincide with the ones obtained in \cite{bds97} for a non-confined 
system in the proper limit, i.e. by making the substitution 
\begin{displaymath}
\Sigma\longrightarrow
\{(\lambda,\sig)\lvert\sig\in\Omega_d\quad\&\quad 0\le\lambda\le 1\}. 
\end{displaymath}

Let us close this section analyzing some properties of the pressure tensor at 
the boundary. From Eq. (\ref{Pijc}), it is clear that, 
for convex borders, 
$P_{ij}^{(c)}(\mathbf{r},t)=0$ if $\mathbf{r}\in\partial V$, because if 
$\mathbf{r}_1(\lambda,\sig)\in V$ then $\mathbf{r}_2(\lambda,\sig)\notin V$ and 
vice versa. Hence, it is $P_{ij}(\mathbf{r},t)=P_{ij}^{(k)}(\mathbf{r},t)$ if 
$\mathbf{r}\in\partial V$. In particular, taking $i=j$ and both in the 
direction of $\mathbf{N}(\mathbf{r})$, the force per unit area that the fluid 
exerts to the wall is identified as 
\begin{equation}\label{pressure}
p(\mathbf{r},t)=\int d\mathbf{v}m[\mathbf{v}\cdot\mathbf{N}(\mathbf{r})]^2
f(\mathbf{r},\mathbf{v},t), 
\quad\textrm{for}\quad \mathbf{r}\in\partial V. 
\end{equation}
This identification can be done because the change in the momentum of any 
particle at the boundary can be only due to the wall-particle force. In 
equilibrium, this result is known as \emph{contact theorem} 
\cite{HansenMcDonald}, where $p_e(\mathbf{r})=n_e(\mathbf{r})T$, with the 
temperature, $T$, being a constant over all the system. Nevertheless, let us 
remark that Eq. (\ref{pressure}) is 
an exact property of any state out of equilibrium. Moreover, it seems that it 
only depends on the interaction of the particles with the wall (the one given 
by Eq. (\ref{opbe})), independently of the interaction between the particles. 

\section{$\mathcal{H}$-theorem}\label{section4}

In this section, it will be shown that the kinetic equation (\ref{enskog}) 
fulfills an $\mathcal{H}$-theorem, i.e. there exists a functional of the 
distribution function, $\mathcal{H}[f]$, such that 
$\frac{d\mathcal{H}[f]}{dt}\leq 0$ for all times and initial conditions. This 
property represents the generalization for physical boundary conditions of 
R\'esibois's result, that was stated for the Enskog equation with periodic 
boundary conditions \cite{rPRL78,r78}. 

Following R\'esibois, the functional $\mathcal{H}$ is chosen to be
\begin{equation}
\mathcal{H}\equiv\int d\Gamma\rho_N(\Gamma,t)\ln\rho_N(\Gamma,t), 
\end{equation}
where $\rho_N(\Gamma,t)$ is taken to be of the form given by (\ref{rhoN}). Let 
us remark that $\rho_N$ is not the actual N-particle distribution of the 
system, but an approximation that can be constructed with the knowledge of 
the one-particle distribution function through Eqs. (\ref{fW}) and (\ref{wF}). 
Then, $\mathcal{H}$ can be expressed in terms of the distribution function, 
obtaining 
\begin{equation}
\mathcal{H}[f]=\mathcal{H}^{(k)}[f]+\mathcal{H}^{(c)}[f], 
\end{equation}
with
\begin{equation}
\mathcal{H}^{(k)}[f]\equiv\int d\mathbf{r}\int d\mathbf{v}
f(\mathbf{r},\mathbf{v},t)[\ln f(\mathbf{r},\mathbf{v},t)-1], 
\end{equation}
the Boltzmann functional, and 
\begin{equation}
\mathcal{H}^{(c)}[f]\equiv-\ln\phi[w]-\int d\mathbf{r}
n(\mathbf{r},t)\ln\frac{n(\mathbf{r},t)}{w(\mathbf{r},t)}, 
\end{equation}
an additional contribution that vanishes in the low-density limit. Note that 
$\mathcal{H}^{(c)}$ is a functional of the density, because $w$ is a functional 
of the density through Eq. (\ref{wF}). 

In Appendix \ref{dh} it is shown that, with the kind of boundary conditions 
being considered here, 
\begin{eqnarray}
\frac{d\mathcal{H}^{(k)}}{dt}=\frac{\sigma^{d-1}}{2}\int d\mathbf{r}\int 
d\mathbf{v}_1\int d\mathbf{v}_2\int_{\Omega(\mathbf{r})}d\sig
\theta(-\mathbf{v}_{12}\cdot\sig)\abs{\mathbf{v}_{12}\cdot\sig}
g_2(\mathbf{r},\mathbf{r}+\boldsymbol{\sigma}\lvert n)\nonumber\\
f(\mathbf{r}+\boldsymbol{\sigma},\mathbf{v}_1,t)f(\mathbf{r},\mathbf{v}_2,t)
\ln\left[\frac{f(\mathbf{r}+\boldsymbol{\sigma},\mathbf{v}_1',t)
f(\mathbf{r},\mathbf{v}_2',t)}
{f(\mathbf{r}+\boldsymbol{\sigma},\mathbf{v}_1,t)
f(\mathbf{r},\mathbf{v}_2,t)}\right]. 
\end{eqnarray}
Employing the inequality
\begin{equation}
x\ln\frac{y}{x}\leq y-x, 
\end{equation}
valid $\forall x,y>0$ and performing standard manipulations, it is obtained 
\begin{equation}\label{dHk}
\frac{d\mathcal{H}^{(k)}}{dt}\leq I(t), 
\end{equation}
where
\begin{equation}
I(t)\equiv\sigma^{d-1}\int d\mathbf{r}\int_{\Omega(\mathbf{r})}d\sig 
g_2(\mathbf{r},\mathbf{r}+\boldsymbol{\sigma}\lvert n)n(\mathbf{r},t)
n(\mathbf{r}+\boldsymbol{\sigma},t)\mathbf{u}(\mathbf{r}+\boldsymbol{\sigma},t)
\cdot\sig. 
\end{equation}
The equality being valid if and only if 
\begin{equation}\label{condition}
f(\mathbf{r},\mathbf{v}_2',t)f(\mathbf{r}+\boldsymbol{\sigma},\mathbf{v}_1',t)=
f(\mathbf{r},\mathbf{v}_2,t)f(\mathbf{r}+\boldsymbol{\sigma},\mathbf{v}_1,t), 
\end{equation}
$\forall\mathbf{r}\in V$, $\forall\sig\in\Omega(\mathbf{r})$ and 
$\forall\mathbf{v}_1, \mathbf{v}_2$ such that $\mathbf{v}_{12}\cdot\sig\leq 0$. 
The time derivative of $\mathcal{H}^{(c)}$ is also calculated in Appendix 
\ref{dh}, obtaining
\begin{equation}
\frac{d\mathcal{H}^{(c)}}{dt}=-I(t),
\end{equation}
so that we can conclude that
\begin{equation}
\frac{d\mathcal{H}}{dt}\leq 0,
\end{equation}
with the equality being valid when the condition given by Eq. (\ref{condition}) 
holds. 

Assuming that the total number of particles and energy are finite, it can be 
shown that $\mathcal{H}$ is bounded from below \cite{resibois}. Hence, if the 
initial 
distribution function, $f(\mathbf{r},\mathbf{v},0)$, is such that $\mathcal{H}$ 
is finite, as $\frac{d\mathcal{H}}{dt}\leq 0$, $\mathcal{H}$ must reach a 
stationary value in the long time limit. This stationary value is only reached 
when $\frac{d\mathcal{H}^{(k)}}{dt}=I$, that means that the distribution 
function must fulfill Eq. (\ref{condition}). Let us label the distribution 
function compatible with a stationary value of $\mathcal{H}$ as $f_0$. By 
taking logarithm in Eq. (\ref{condition}), it is obtained

\begin{equation}\label{conditionLn}
\ln f_0(\mathbf{r},\mathbf{v}_2',t)
+\ln f_0(\mathbf{r}+\boldsymbol{\sigma},\mathbf{v}_1',t)=
\ln f_0(\mathbf{r},\mathbf{v}_2,t)
+\ln f_0(\mathbf{r}+\boldsymbol{\sigma},\mathbf{v}_1,t), 
\end{equation}
$\forall\mathbf{r}\in V$, $\forall\sig\in\Omega(\mathbf{r})$ and 
$\forall\mathbf{v}_1, \mathbf{v}_2$, where the restriction 
$\mathbf{v}_{12}\cdot\sig\leq 0$ has been eliminated because 
$\mathbf{v}_{12}'\cdot\sig\geq 0$ and $(\mathbf{v}_i')'=\mathbf{v}_i$. Eq. 
(\ref{conditionLn}) implies that $\ln f_0$ must be a quantity that is conserved 
in a binary collision, usually called ``collision invariant''. The most 
general collision invariant in a binary collision is a linear combination of 
the number of particles, total linear momentum, total energy and total 
angular momentum \cite{landau}. Therefore, $\ln f_0$ must be of the form 
\begin{equation}
\ln f_0(\mathbf{r},\mathbf{v},t)=A_0(\mathbf{r},t)+\mathbf{A}_1(t)\cdot
\mathbf{v}+A_2(t)v^2+\mathbf{A}_3(t)\cdot(\mathbf{r}\times\mathbf{v}). 
\end{equation}
Equivalently, the distribution can be written in the form
\begin{equation}\label{fe0}
f_0(\mathbf{r},\mathbf{v},t)=n(\mathbf{r},t)
\chi_M[\mathbf{v}-\mathbf{u}(\mathbf{r},t),T(t)], 
\end{equation}
where $\chi_M$ is the Maxwellian distribution introduced in Eq. (\ref{max}), 
and $n$, $\mathbf{u}$ and $T$ can be interpreted as the corresponding density, 
flow velocity, and temperature associated to $f_0$. As $\mathbf{A}_1$, $A_2$ 
and $\mathbf{A}_3$ are arbitrary functions of time but do not depend on 
position, it can be concluded that $T$ is an 
arbitrary function  of time and $\mathbf{u}$ is of the form 
\begin{equation}\label{u}
\mathbf{u}(\mathbf{r},t)=\mathbf{u}_0(t)+\mathbf{w}(t)\times\mathbf{r}, 
\end{equation}
i.e. a translation plus a rotation. Moreover, $n$ is an arbitrary function of 
the position and time. 
Applying the boundary conditions, it is 
concluded that, in general, $\mathbf{u}(\mathbf{r},t)=\mathbf{0}$. In effect, 
if $\mathbf{u}$ is of the form given by Eq. (\ref{u}), and 
$\mathbf{N}(\mathbf{r})\cdot\mathbf{u}(\mathbf{r},t)=0, 
\forall\mathbf{r}\in\partial V$ and $\forall t$, then 
$\mathbf{u}(\mathbf{r},t)=\mathbf{0}, \forall\mathbf{r}\in V$ and $\forall t$. 
However, for some particular geometries, there can be exceptions. For example, 
$\mathbf{u}(\mathbf{r},t)=\mathbf{w}(t)\times\mathbf{r}$ is compatible with a 
circular shape in $d=2$,  with an spherical volume in $d=3$, or with a 
cylinder if its axes is in the direction of $\mathbf{w}$. Moreover, 
taking into account the continuity equation, Eq. (\ref{eh1}), we obtain that 
$\frac{\partial n(\mathbf{r},t)}{\partial t}=0$ and 
\begin{equation}\label{fe2}
f_0(\mathbf{r},\mathbf{v},t)=n(\mathbf{r})
\chi_M[\mathbf{v},T(t)]. 
\end{equation}
As the total energy is time-independent, $T$ is also time-independent. Finally, 
by substituting Eq. (\ref{fe2}) with a time-independent temperature into 
the Enskog equation, it is obtained that the function $n$ 
satisfies the same equation that $n_e$, Eq. (\ref{eqne}). If we consider 
situations for which it has only one 
solution, it is concluded that, for the considered initial conditions, 
$f(\mathbf{r},\mathbf{v},t)\to f_0(\mathbf{r},\mathbf{v})\equiv 
f_e(\mathbf{r},\mathbf{v})$ in the long time limit. 

\section{Conclusions}\label{section5}

In this paper, we have formulated a kinetic equation that describes the 
dynamics of a system composed of elastic hard spheres or disks confined with 
an arbitrary hard wall (also elastic). The equation is derived under the same 
hypothesis used to derive the MEE and its range of validity is supposed to be 
the same. In the bulk, the obtained equation coincides 
with the MEE but, closed to the boundary, the collision operator changes, and 
takes into account that only some collisions are possible, due to the 
geometrical constraints imposed by the boundary. Let us note that the equation 
can be easily 
generalized to incorporate other collision rules (as, for example, inelastic 
collisions \cite{c90,bds97,brs13} or models of active matter \cite{metal13}), 
by slightly modifying the collision operator, $J_E$. In 
the same lines, other kind of collisions with the confining wall may be 
considered by modifying the boundary conditions of the one-particle 
distribution function, $f$. The important ingredient for the derivation is that 
the particles are hard spheres or disks and the wall is hard. In addition, a 
simplified equation is derived that is supposed to be valid for densities 
between Boltzmann and Enskog and that works remarkably well in the monolayer 
case \cite{bmg16,bgm17}. 

From the kinetic equation, balance equations for the hydrodynamic fields are 
derived. These are the starting point for a subsequent derivation of the 
hydrodynamic equations, for example, via the Chapman-Enskog method. As in the 
MEE, the fluxes can be decomposed in a kinetic part plus a 
collisional transfer contribution. Closed to the boundary, this later 
contribution is different from the one derived from the MEE (again, due to 
geometrical constraints) and this may imply the need to modify in a 
non-trivial way 
the structure of hydrodynamics. In this sense, the analysis made here opens the 
possibility of exploring the form of the hydrodynamic equations close to 
the boundary, and it can help to study instabilities in shaken granular fluids 
\cite{mvprkeu05,ris07,rpgrsc11,cms12}, that are still not well understood 
although they seem to have a hydrodynamic character \cite{ka11}. 

Finally, we have shown that the kinetic equation admits an 
$\mathcal{H}$-theorem. Using the same functional as R\'esibois took for the 
bulk 
MEE, it has been proved that $\frac{d\mathcal{H}}{dt}\le 0$ for any solution of 
the kinetic equation. Moreover, it is shown that, in the long time limit, the 
system reaches the known inhomogeneous equilibrium distribution function: a 
Maxwellian distribution with a constant temperature and the proper density 
profile given by Statistical Mechanics. In our opinion, the result is 
remarkable because, despite the approximate character of the kinetic equation, 
it demonstrates the approach to equilibrium of the one-particle distribution 
function of a strong interacting system with a finite number of degrees of 
freedom and with realistic boundary conditions. Let us note that the limitation 
of R\'esibois result to periodic boundary conditions is mentioned several times 
in the literature \cite{rPRL78,r78,p87}; in this context, it is seen that the 
solution to this limitation resides in the correct extension of the MEE to 
incorporate the boundary consistently.

\begin{acknowledgements}
This research was supported by the Ministerio de Educaci\'{o}n y
Ciencia (Spain) through Grant No. FIS2014-53808-P (partially financed
by FEDER funds). 
\end{acknowledgements}

\appendix

\section{Evaluation of the collisional fluxes}\label{cf}
The objective of this appendix is to evaluate the collisional contribution to 
the pressure tensor and heat flux. We will proceed using intuitive arguments, 
taking into account the collisions that contribute to the flux with their 
corresponding momentum or energy interchange. 

\begin{figure}
\begin{center}
\includegraphics[angle=0,width=0.5\linewidth,clip]{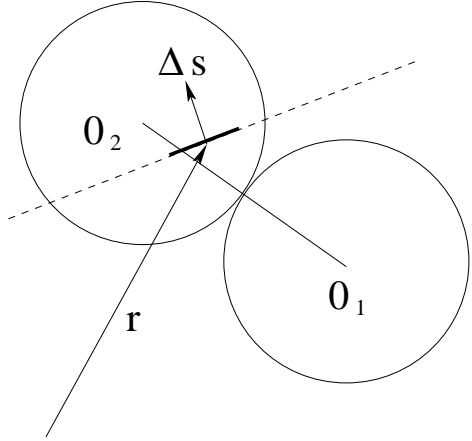}
\end{center}
\caption{Sketch of a typical collision that contributes to the flux through 
$\Delta\mathbf{s}$. It is assumed that particles $1$ and $2$ 
have the centers at $0_1$ and $0_2$ respectively. }
\label{fap}
\end{figure}

Let us first analyze the pressure tensor case. Let us consider a surface 
element, $\Delta\mathbf{s}$, centered at $\mathbf{r}$ and two particles at 
contact in such a way that the line joining the two centers cross the surface 
(see Fig. \ref{fap}). When the collision takes place, the variation of the 
momentum of particle $2$ is 
\begin{equation}\label{deltap}
\Delta p_{2,i}=m(\sig\cdot\mathbf{v}_{12})\hat{\sigma}_i.
\end{equation}

It is assumed that, 
in order to evaluate the flux, $\Delta p_{2,i}$ cross the surface through the 
intersection of the surface with the line joining the two particles. To 
calculate the collisional contribution to the flux, we have to consider all 
the possible collisions of this kind with its corresponding $\Delta p_{2,i}$. 
The surface divides the space in two regions; of course, the centers of the 
particles must be in different regions. We will consider that particle $2$ is 
in the region pointed by $\Delta\mathbf{s}$, as in the Figure. The center of 
particle $1$ can be parameterized by 
\begin{equation}\label{r1}
\mathbf{r}_1(\lambda, \sig)=\mathbf{r}+\lambda\sigma\sig, 
\end{equation}
with $\lambda\in(0,1)$ and $\sig$ a unitary vector of arbitrary orientation, 
but compatible with $\Delta\mathbf{s}$, i.e. $\sig\cdot\Delta\mathbf{s}<0$. In 
these conditions, particle $2$ must be in a solid angle
\begin{equation}
\Delta\hat{\sigma}_2=\frac{\abs{\sig\cdot\Delta\mathbf{s}}}
{(\lambda\sigma)^{d-1}}, 
\end{equation}
around
\begin{equation}
\mathbf{r}_2(\lambda, \sig)=\mathbf{r}-(1-\lambda)\sigma\sig. 
\end{equation}
Note that we have used the same notation for the $\sig$ of the collision in Eq. 
(\ref{deltap}), and for the parameter to specify the position of particle $1$ 
in Eq. (\ref{r1}). This can be done because its difference is of order 
$\Delta\hat{\sigma}_2$. 

Let us consider that particle $1$ is in the volume element 
$\Delta\mathbf{r}_1=(\lambda\sigma)^{d-1}\Delta(\lambda\sigma)\Delta\sig$ 
parameterized by $(\lambda,\sig)$. Hence, if particle $2$ collides with 
particle 
$1$ in the time interval $\Delta t$ with $\sig$, it is in the volume element 
$\Delta\mathbf{r}_2=\sigma^{d-1}\Delta\hat{\sigma}_2
\abs{\mathbf{v}_{12}\cdot\sig}\Delta t$. Then, the total number of collisions 
that contribute to the flux for given $\mathbf{v}_1$ and $\mathbf{v}_2$ is
\begin{eqnarray}
\theta(-\mathbf{v}_{12}\cdot\sig)f_2[\mathbf{r}_1(\lambda,\sig),\mathbf{v}_1,
\mathbf{r}_2(\lambda,\sig),\mathbf{v}_2]\Delta\mathbf{r}_1\Delta\mathbf{v}_1
\Delta\mathbf{r}_2\Delta\mathbf{v}_2\nonumber\\
=\sigma^d\theta(-\mathbf{v}_{12}\cdot\sig)
f_2[\mathbf{r}_1(\lambda,\sig),\mathbf{v}_1,
\mathbf{r}_2(\lambda,\sig),\mathbf{v}_2]\abs{\sig\cdot\Delta\mathbf{s}}
\abs{\mathbf{v}_{12}\cdot\sig}\Delta\mathbf{v}_1\Delta\mathbf{v}_2\Delta\sig
\Delta\lambda\Delta t. 
\end{eqnarray}
Let us take $\Delta\mathbf{s}=\Delta s\mathbf{e}_j$ where $\mathbf{e}_j$ 
is a unit vector in the direction of one of our coordinate axes. The amount 
of momentum that travels through the surface in the direction of 
$\Delta\mathbf{s}$ per unit time and area due to collisions of particles with 
velocities $\mathbf{v}_1$ and $\mathbf{v}_2$ is then
\begin{equation}\label{deltapc}
\Delta P_{ij}^{(c)}=m\sigma^d
\theta(-\mathbf{v}_{12}\cdot\sig)
f_2[\mathbf{r}_1(\lambda,\sig),\mathbf{v}_1,
\mathbf{r}_2(\lambda,\sig),\mathbf{v}_2]\hat{\sigma}_i\hat{\sigma}_j
(\mathbf{v}_{12}\cdot\sig)^2\Delta\mathbf{v}_1\Delta\mathbf{v}_2\Delta\sig
\Delta\lambda,  
\end{equation}
where we have taken into account that 
$\abs{\mathbf{v}_{12}\cdot\sig}=-\mathbf{v}_{12}\cdot\sig$ and 
$\abs{\hat{\sigma}_j}=-\hat{\sigma}_j$. 
The net collisional pressure tensor is obtained integrating in Eq. 
(\ref{deltapc}) for all the allowed 
collisions. 

Far from the boundary, when there are not geometrical constraints, 
the result is
\begin{equation}\label{pc1}
P_{ij}^{(c)}=m\sigma^d\int d\mathbf{v}_1\int d\mathbf{v}_2\int_0^1d\lambda
\int_{\hat{\sigma}_j<0} d\sig
\theta(-\mathbf{v}_{12}\cdot\sig)
f_2[\mathbf{r}_1(\lambda,\sig),\mathbf{v}_1,
\mathbf{r}_2(\lambda,\sig),\mathbf{v}_2]\hat{\sigma}_i\hat{\sigma}_j
(\mathbf{v}_{12}\cdot\sig)^2. 
\end{equation}
The integration can also be done summing for $\hat{\sigma}_j>0$ but, then, the 
amount of momentum that crosses the surface is $\Delta p_{1,i}=-\Delta p_{2,i}$, 
so that
\begin{equation}
P_{ij}^{(c)}=m\sigma^d\int d\mathbf{v}_1\int d\mathbf{v}_2\int_0^1d\lambda
\int_{\hat{\sigma}_j>0} d\sig
\theta(-\mathbf{v}_{12}\cdot\sig)
f_2[\mathbf{r}_1(\lambda,\sig),\mathbf{v}_1,
\mathbf{r}_2(\lambda,\sig),\mathbf{v}_2]\hat{\sigma}_i\hat{\sigma}_j
(\mathbf{v}_{12}\cdot\sig)^2, 
\end{equation}
because, in this case, $\abs{\hat{\sigma}_j}=\hat{\sigma}_j$. Hence, we can 
re-write Eq. (\ref{pc1}) as
\begin{equation}
P_{ij}^{(c)}=\frac{m}{2}\sigma^d
\int d\mathbf{v}_1\int d\mathbf{v}_2\int_0^1d\lambda
\int d\sig
\theta(-\mathbf{v}_{12}\cdot\sig)
f_2[\mathbf{r}_1(\lambda,\sig),\mathbf{v}_1,
\mathbf{r}_2(\lambda,\sig),\mathbf{v}_2]\hat{\sigma}_i\hat{\sigma}_j
(\mathbf{v}_{12}\cdot\sig)^2, 
\end{equation}
that coincides with the expression derived in \cite{bds97} for $d=3$ when the 
factorization for $f_2$ given by Eq. (\ref{f2}) is used. 

If there are geometrical constraints, we proceed similarly. Integrating in 
Eq. (\ref{deltapc}) for the allowed collisions, it is obtained 
\begin{equation}\label{pcc1}
P_{ij}^{(c)}=m\sigma^d\int d\mathbf{v}_1\int d\mathbf{v}_2\int\int_{\Sigma^-} 
d\lambda d\sig
\theta(-\mathbf{v}_{12}\cdot\sig)
f_2[\mathbf{r}_1(\lambda,\sig),\mathbf{v}_1,
\mathbf{r}_2(\lambda,\sig),\mathbf{v}_2]\hat{\sigma}_i\hat{\sigma}_j
(\mathbf{v}_{12}\cdot\sig)^2, 
\end{equation}
where the region of integration in $(\lambda, \sig)$ is
\begin{equation}
\Sigma^-=\{(\lambda,\sig)\lvert\sig\in\Omega_d\quad\textrm{with}\quad
\hat{\sigma}_j<0
\quad\&\quad 0\le\lambda\le 1
\quad\&\quad\mathbf{r}_1(\lambda,\sig)\in V
\quad\&\quad\mathbf{r}_2(\lambda,\sig)\in V \}. 
\end{equation}
$P_{ij}^{(c)}$ can also be calculated summing for $\hat{\sigma}_j>0$ but, then, 
the amount of momentum that crosses the surface is 
$\Delta p_{1,i}=-\Delta p_{2,i}$, so that
\begin{equation}
P_{ij}^{(c)}=m\sigma^d\int d\mathbf{v}_1\int d\mathbf{v}_2\int\int_{\Sigma^+} 
d\lambda d\sig
\theta(-\mathbf{v}_{12}\cdot\sig)
f_2[\mathbf{r}_1(\lambda,\sig),\mathbf{v}_1,
\mathbf{r}_2(\lambda,\sig),\mathbf{v}_2]\hat{\sigma}_i\hat{\sigma}_j
(\mathbf{v}_{12}\cdot\sig)^2, 
\end{equation}
where the region of integration in $(\lambda, \sig)$ is
\begin{equation}
\Sigma^+=\{(\lambda,\sig)\lvert\sig\in\Omega_d\quad\textrm{with}\quad
\hat{\sigma}_j>0
\quad\&\quad 0\le\lambda\le 1
\quad\&\quad\mathbf{r}_1(\lambda,\sig)\in V
\quad\&\quad\mathbf{r}_2(\lambda,\sig)\in V \}. 
\end{equation}
Hence, we can re-write Eq. (\ref{pcc1}) as
\begin{equation}
P_{ij}^{(c)}=\frac{m}{2}
\sigma^d\int d\mathbf{v}_1\int d\mathbf{v}_2\int\int_{\Sigma} 
d\lambda d\sig
\theta(-\mathbf{v}_{12}\cdot\sig)
f_2[\mathbf{r}_1(\lambda,\sig),\mathbf{v}_1,
\mathbf{r}_2(\lambda,\sig),\mathbf{v}_2]\hat{\sigma}_i\hat{\sigma}_j
(\mathbf{v}_{12}\cdot\sig)^2, 
\end{equation}
where the region of integration in $(\lambda, \sig)$ is
\begin{equation}
\Sigma=\{(\lambda,\sig)\lvert\sig\in\Omega_d\quad\&\quad 0\le\lambda\le 1
\quad\&\quad\mathbf{r}_1(\lambda,\sig)\in V
\quad\&\quad\mathbf{r}_2(\lambda,\sig)\in V \}. 
\end{equation}

To calculate the collisional contribution to the energy flux, $J_{E,j}^{(c)}$, 
the analysis is 
similar, but taking into account that, when the collision 
takes place, the variation of the energy of particle $2$ is 
\begin{equation}\label{deltae}
\Delta e_{2,i}=\frac{m}{2}(\sig\cdot\mathbf{v}_{12})^2
+m(\sig\cdot\mathbf{v}_{12})(\sig\cdot\mathbf{v}_{2}). 
\end{equation}
Once $J_{E,j}^{(c)}$ is calculated, the heat flux is expressed as 
$q_{j}^{(c)}=J_{E,j}^{(c)}-\sum_{i}u_iP_{ij}^{(c)}$. 

\section{Evaluation of the divergence of the collisional fluxes}
\label{dcf}
As in the previous Appendix, we focus on the pressure tensor because the heat 
flux case is similar. Let us first re-write the collisional pressure tensor 
given by Eq. (\ref{Pijc}) in the form
\begin{equation}\label{Pijca}
P_{ij}^{(c)}(\mathbf{r},t)
=\frac{m}{2}\sigma^d\int d\mathbf{v}_1\int d\mathbf{v}_2\int
d\sig\int_{\lambda_1(\mathbf{r},\sig)}^{\lambda_2(\mathbf{r},\sig)}
d\lambda 
\theta(-\mathbf{v}_{12}\cdot\sig)f_2[\mathbf{r}_1(\lambda,\sig),
\mathbf{v}_1,\mathbf{r}_2(\lambda,\sig),\mathbf{v}_2,t]
(\mathbf{v}_{12}\cdot\sig)^2\hat{\sigma_i}\hat{\sigma_j}, 
\end{equation}
where $\lambda_1(\mathbf{r},\sig)$ and $\lambda_2(\mathbf{r},\sig)$ are such 
that $\sigma\lambda_1(\mathbf{r},\sig)$ and $\sigma\lambda_2(\mathbf{r},\sig)$ 
are the minimum and maximum distance from $\mathbf{r}$ to 
$\mathbf{r}_1(\lambda,\sig)$ respectively, for a given orientation, $\sig$. 
In the bulk of the system, we trivially have $\lambda_1(\mathbf{r},\sig)=0$ 
and $\lambda_2(\mathbf{r},\sig)=1$, for all $\sig$, but closed to the boundary 
these functions depend on the geometry of it. 

Taking into account Eq. (\ref{Pijca}), the divergence of $P_{ij}^{(c)}$ can be 
expressed as 
\begin{eqnarray}\label{dpc}
&&\frac{\partial}{\partial\mathbf{r}}\cdot P_{ij}^{(c)}(\mathbf{r})\nonumber\\
&&=\frac{m}{2}\sigma^d\int d\mathbf{v}_1\int d\mathbf{v}_2\int d\sig 
\theta(-\mathbf{v}_{12}\cdot\sig)(\mathbf{v}_{12}\cdot\sig)^2\sig\sig\cdot
\frac{\partial}{\partial\mathbf{r}}\lambda_2
f_2[\mathbf{r}_1(\lambda_2,\sig),
\mathbf{v}_1,\mathbf{r}_2(\lambda_2,\sig),\mathbf{v}_2]
\nonumber\\
&&-\frac{m}{2}\sigma^d\int d\mathbf{v}_1\int d\mathbf{v}_2\int d\sig 
\theta(-\mathbf{v}_{12}\cdot\sig)(\mathbf{v}_{12}\cdot\sig)^2\sig\sig\cdot
\frac{\partial}{\partial\mathbf{r}}\lambda_1
f_2[\mathbf{r}_1(\lambda_1,\sig),
\mathbf{v}_1,\mathbf{r}_2(\lambda_1,\sig),\mathbf{v}_2]\nonumber\\
&&+\frac{m}{2}\sigma^d\int d\mathbf{v}_1\int d\mathbf{v}_2\int d\sig 
\theta(-\mathbf{v}_{12}\cdot\sig)(\mathbf{v}_{12}\cdot\sig)^2\sig
\int_{\lambda_1(\mathbf{r},\sig)}^{\lambda_2(\mathbf{r},\sig)} d\lambda
\sig\cdot\frac{\partial}{\partial\mathbf{r}}
f_2[\mathbf{r}_1(\lambda,\sig),
\mathbf{v}_1,\mathbf{r}_2(\lambda,\sig),\mathbf{v}_2]. \nonumber\\
\end{eqnarray}
Taking into account that 
\begin{equation}
\frac{\partial}{\partial\lambda}f_2[\mathbf{r}_1(\lambda,\sig),\mathbf{v}_1,
\mathbf{r}_2(\lambda,\sig),\mathbf{v}_2]=\boldsymbol{\sigma}\cdot
\frac{\partial}{\partial\mathbf{r}}
f_2[\mathbf{r}_1(\lambda,\sig),\mathbf{v}_1,
\mathbf{r}_2(\lambda,\sig),\mathbf{v}_2], 
\end{equation}
the last term of the r.h.s. of Eq. (\ref{dpc}) can be written as
\begin{eqnarray}\label{mo1}
\frac{m}{2}\sigma^d\int d\mathbf{v}_1\int d\mathbf{v}_2\int d\sig 
\theta(-\mathbf{v}_{12}\cdot\sig)(\mathbf{v}_{12}\cdot\sig)^2\sig
\int_{\lambda_1(\mathbf{r},\sig)}^{\lambda_2(\mathbf{r},\sig)} d\lambda
\sig\cdot\frac{\partial}{\partial\mathbf{r}}
f_2[\mathbf{r}_1(\lambda,\sig),
\mathbf{v}_1,\mathbf{r}_2(\lambda,\sig),\mathbf{v}_2]\nonumber\\
=\frac{m}{2}\sigma^{d-1}\int d\mathbf{v}_1\int d\mathbf{v}_2\int d\sig 
\theta(-\mathbf{v}_{12}\cdot\sig)(\mathbf{v}_{12}\cdot\sig)^2\sig
f_2[\mathbf{r}_1(\lambda_2,\sig),
\mathbf{v}_1,\mathbf{r}_2(\lambda_2,\sig),\mathbf{v}_2]\nonumber\\
-\frac{m}{2}\sigma^{d-1}\int d\mathbf{v}_1\int d\mathbf{v}_2\int d\sig 
\theta(-\mathbf{v}_{12}\cdot\sig)(\mathbf{v}_{12}\cdot\sig)^2\sig
f_2[\mathbf{r}_1(\lambda_1,\sig),
\mathbf{v}_1,\mathbf{r}_2(\lambda_1,\sig),\mathbf{v}_2]. \nonumber\\
\end{eqnarray}
Changing variables, 
\begin{eqnarray}
\mathbf{v}_1&\leftrightarrow&\mathbf{v}_2, \\
\sig&\to&-\sig, 
\end{eqnarray}
in the second term of the r.h.s., it is obtained
\begin{eqnarray}
\frac{m}{2}\sigma^{d-1}\int d\mathbf{v}_1\int d\mathbf{v}_2\int d\sig 
\theta(-\mathbf{v}_{12}\cdot\sig)(\mathbf{v}_{12}\cdot\sig)^2\sig
f_2[\mathbf{r}_1(\lambda_1,\sig),
\mathbf{v}_1,\mathbf{r}_2(\lambda_1,\sig),\mathbf{v}_2]. \nonumber\\
=-\frac{m}{2}\sigma^{d-1}\int d\mathbf{v}_1\int d\mathbf{v}_2\int d\sig 
\theta(-\mathbf{v}_{12}\cdot\sig)(\mathbf{v}_{12}\cdot\sig)^2\sig
f_2[\mathbf{r}_1(\lambda_2,\sig),
\mathbf{v}_1,\mathbf{r}_2(\lambda_2,\sig),\mathbf{v}_2],  \nonumber\\
\end{eqnarray}
where it has taken into account that 
\begin{eqnarray}
\mathbf{r}_1[\lambda_1(\mathbf{r},-\sig),-\sig]
&=&\mathbf{r}_2[\lambda_2(\mathbf{r},\sig),\sig], \\
\mathbf{r}_2[\lambda_1(\mathbf{r},-\sig),-\sig]
&=&\mathbf{r}_1[\lambda_2(\mathbf{r},\sig),\sig]. 
\end{eqnarray}
So, we have 
\begin{eqnarray}\label{mo2}
\frac{m}{2}\sigma^d\int d\mathbf{v}_1\int d\mathbf{v}_2\int d\sig 
\theta(-\mathbf{v}_{12}\cdot\sig)(\mathbf{v}_{12}\cdot\sig)^2\sig
\int_{\lambda_1(\mathbf{r},\sig)}^{\lambda_2(\mathbf{r},\sig)} d\lambda
\sig\cdot\frac{\partial}{\partial\mathbf{r}}
f_2[\mathbf{r}_1(\lambda,\sig),
\mathbf{v}_1,\mathbf{r}_2(\lambda,\sig),\mathbf{v}_2]\nonumber\\
=m\sigma^{d-1}\int d\mathbf{v}_1\int d\mathbf{v}_2\int d\sig 
\theta(-\mathbf{v}_{12}\cdot\sig)(\mathbf{v}_{12}\cdot\sig)^2\sig
f_2[\mathbf{r}_1(\lambda_2,\sig),
\mathbf{v}_1,\mathbf{r}_2(\lambda_2,\sig),\mathbf{v}_2]. \nonumber\\
\end{eqnarray}
Performing the same change of variables in the second term of the r.h.s. of 
Eq. (\ref{dpc}), it is obtained
\begin{eqnarray}\label{mo3}
\frac{m}{2}\sigma^d\int d\mathbf{v}_1\int d\mathbf{v}_2\int d\sig 
\theta(-\mathbf{v}_{12}\cdot\sig)(\mathbf{v}_{12}\cdot\sig)^2\sig\sig\cdot
\frac{\partial}{\partial\mathbf{r}}\lambda_1
f_2[\mathbf{r}_1(\lambda_1,\sig),
\mathbf{v}_1,\mathbf{r}_2(\lambda_1,\sig),\mathbf{v}_2]\nonumber\\
=-\frac{m}{2}\sigma^d\int d\mathbf{v}_1\int d\mathbf{v}_2\int d\sig 
\theta(-\mathbf{v}_{12}\cdot\sig)(\mathbf{v}_{12}\cdot\sig)^2\sig\sig\cdot
\frac{\partial}{\partial\mathbf{r}}\lambda_2
f_2[\mathbf{r}_1(\lambda_2,\sig),
\mathbf{v}_1,\mathbf{r}_2(\lambda_2,\sig),\mathbf{v}_2].  \nonumber\\
\end{eqnarray}
By substituting Eqs. (\ref{mo2}) and (\ref{mo3}) into Eq. (\ref{dpc}), 
it is obtained
\begin{eqnarray}\label{pijca2}
&&\frac{\partial}{\partial\mathbf{r}}\cdot P_{ij}^{(c)}(\mathbf{r})\nonumber\\
\nonumber\\
&&=m\sigma^{d-1}\int d\mathbf{v}_1\int d\mathbf{v}_2\int d\sig 
\theta(-\mathbf{v}_{12}\cdot\sig)(\mathbf{v}_{12}\cdot\sig)^2\sig
\left[1+\boldsymbol{\sigma}\cdot\frac{\partial}{\partial\mathbf{r}}
\lambda_2\right]
f_2[\mathbf{r}_1(\lambda_2,\sig),
\mathbf{v}_1,\mathbf{r}_2(\lambda_2,\sig),\mathbf{v}_2]. \nonumber\\
\end{eqnarray}

Now, let us analyze the function 
$1+\boldsymbol{\sigma}\cdot\frac{\partial}{\partial\mathbf{r}}\lambda_2$. Let 
us first consider the simplest case of a plane located at $z=-\sigma/2$. If 
$\sig\in\Omega(\mathbf{r})$, then $\lambda_2(\mathbf{r},\sig)=1$. Let us 
define $\Omega^+(\mathbf{r})$, such that
\begin{equation}
\Omega(\mathbf{r})\cup\Omega^+(\mathbf{r})
=\Omega_d. 
\end{equation}
For a given $\mathbf{r}$, it is
\begin{equation}
z=-\lambda_2(\mathbf{r},\sig)\sigma\hat{\sigma}_z, \quad\textrm{for}\quad
\sig\in\Omega^+(\mathbf{r}), 
\end{equation}
so that, for this simple case, we have 
\begin{equation}\label{unolambda}
1+\boldsymbol{\sigma}\cdot\frac{\partial}{\partial\mathbf{r}}
\lambda_2(\mathbf{r},\sig)
=\left\{ \begin{array}{l}
1\quad\textrm{if $\sig\in\Omega(\mathbf{r})$}\\
0\quad\textrm{if $\sig\in\Omega^+(\mathbf{r})$}
\end{array} \right.
\end{equation}
In fact, if the plane has a different orientation, the result is the same 
because the function is a scalar. Moreover, in the general case of an 
arbitrary wall, the result also holds if the tangent plane is defined at 
$\mathbf{r}+\lambda_2\sigma\sig$. 

Hence, by substituting Eq. (\ref{unolambda}) into Eq. (\ref{pijca2}), it is 
finally obtained
\begin{eqnarray}\label{pijcfinal}
&&\frac{\partial}{\partial\mathbf{r}}\cdot P_{ij}^{(c)}(\mathbf{r})\nonumber\\
\nonumber\\
&&=m\sigma^{d-1}\int d\mathbf{v}_1\int d\mathbf{v}_2\int_{\Omega(\mathbf{r})} d\sig 
\theta(-\mathbf{v}_{12}\cdot\sig)(\mathbf{v}_{12}\cdot\sig)^2\sig
f_2(\mathbf{r}+\boldsymbol{\sigma},
\mathbf{v}_1,\mathbf{r},\mathbf{v}_2),   \nonumber\\
\end{eqnarray}
where it has been used that $\lambda_2(\mathbf{r},\sig)=1$ if 
$\sig\in\Omega(\mathbf{r})$.

It still remains to show that 
$\frac{\partial}{\partial\mathbf{r}}\cdot P_{ij}^{(c)}$ coincides with 
$\int d\mathbf{v}m\mathbf{v}J[f_2]$. By standard manipulations, it can be 
shown that 
\begin{equation}\label{psiJ}
\int d\mathbf{v}\psi(\mathbf{v})J[f_2]=\sigma^{d-1}\int d\mathbf{v}_1\int 
d\mathbf{v}_2\int_{\Omega(\mathbf{r})} d\sig\theta(-\mathbf{v}_{12}\cdot\sig)
\abs{\mathbf{v}_{12}\cdot\sig}
f_2(\mathbf{r}+\boldsymbol{\sigma},\mathbf{v}_1,\mathbf{r},\mathbf{v}_2)
(b_{\sig}-1)\psi(\mathbf{v}_2). 
\end{equation}
Taking $\psi(\mathbf{v})=\mathbf{v}_i$, it is  
\begin{equation}\label{mie}
\int d\mathbf{v}\mathbf{v}J[f_2]=
-\sigma^{d-1}\int d\mathbf{v}_1\int d\mathbf{v}_2\int_{\Omega(\mathbf{r})} d\sig 
\theta(-\mathbf{v}_{12}\cdot\sig)(\mathbf{v}_{12}\cdot\sig)^2\sig
f_2(\mathbf{r}+\boldsymbol{\sigma},
\mathbf{v}_1,\mathbf{r},\mathbf{v}_2).  
\end{equation}
Comparing Eq. (\ref{mie}) with Eq. (\ref{pijcfinal}), we finally have 
\begin{equation}
\int d\mathbf{v}m\mathbf{v}J[f_2]=
-\frac{\partial}{\partial\mathbf{r}}\cdot P^{(c)}, 
\end{equation}
as we wanted to prove.


\section{Evaluation of the time derivative of $\mathcal{H}$ }
\label{dh}

Let us first calculate $\frac{d\mathcal{H}^{(k)}}{dt}$. Using standard 
manipulations and applying the boundary conditions, it is obtained 
\begin{equation}
\frac{d\mathcal{H}^{(k)}}{dt}=\int d\mathbf{r}\int d\mathbf{v}J_E[f\lvert f]
\ln f(\mathbf{r},\mathbf{v},t). 
\end{equation}
Eq. (\ref{psiJ}) reduces in the Enskog case to 
\begin{eqnarray}
\int d\mathbf{v}\psi(\mathbf{v})J_E[f\lvert f]
=\sigma^{d-1}\int d\mathbf{v}_1\int d\mathbf{v}_2
\int_{\Omega(\mathbf{r})} d\sig\theta(-\mathbf{v}_{12}\cdot\sig)
\abs{\mathbf{v}_{12}\cdot\sig}
g_2(\mathbf{r}+\boldsymbol{\sigma},\mathbf{r}\lvert n)\nonumber\\
f(\mathbf{r}+\boldsymbol{\sigma},\mathbf{v}_1,t)
f(\mathbf{r},\mathbf{v}_2,t)
(b_{\sig}-1)\psi(\mathbf{v}_2). 
\end{eqnarray}
By taking $\psi=\ln f$, we have
\begin{eqnarray}\label{dhk1}
\frac{d\mathcal{H}^{(k)}}{dt}=\sigma^{d-1}\int  d\mathbf{r}\int d\mathbf{v}_1
\int d\mathbf{v}_2\int_{\Omega(\mathbf{r})}d\sig\theta(-\mathbf{v}_{12}\cdot\sig)
\abs{\mathbf{v}_{12}\cdot\sig}
g_2(\mathbf{r}+\boldsymbol{\sigma},\mathbf{r}\lvert n)\nonumber\\
f(\mathbf{r}+\boldsymbol{\sigma},\mathbf{v}_1,t)
f(\mathbf{r},\mathbf{v}_2,t)
\ln\frac{f(\mathbf{r},\mathbf{v}_2',t)}{f(\mathbf{r},\mathbf{v}_2,t)}. 
\end{eqnarray}
Changing variables, 
\begin{eqnarray}
\mathbf{v}_1&\leftrightarrow&\mathbf{v}_2, \\
\sig&\to&-\sig, 
\end{eqnarray}
Eq. (\ref{dhk1}) is transformed into 
\begin{eqnarray}\label{dhk2}
\frac{d\mathcal{H}^{(k)}}{dt}=\sigma^{d-1}\int d\mathbf{r}\int d\mathbf{v}_1
\int d\mathbf{v}_2\int_{\widetilde{\Omega}(\mathbf{r})}d\sig
\theta(-\mathbf{v}_{12}\cdot\sig)
\abs{\mathbf{v}_{12}\cdot\sig}
g_2(\mathbf{r}-\boldsymbol{\sigma},\mathbf{r}\lvert n)\nonumber\\
f(\mathbf{r}-\boldsymbol{\sigma},\mathbf{v}_2,t)
f(\mathbf{r},\mathbf{v}_1,t)
\ln\frac{f(\mathbf{r},\mathbf{v}_1',t)}{f(\mathbf{r},\mathbf{v}_1,t)}, 
\end{eqnarray}
where, now, the angular integration is taken over the new region, 
$\widetilde{\Omega}(\mathbf{r})$, defined in such a way that 
$\sig\in\widetilde{\Omega}(\mathbf{r})$ if and only if 
$\mathbf{r}-\boldsymbol{\sigma}\in V$. Finally, by changing the space 
variable, $\mathbf{r}\to\mathbf{r}+\boldsymbol{\sigma}$, it is obtained 
\begin{eqnarray}\label{dhk3}
\frac{d\mathcal{H}^{(k)}}{dt}=\sigma^{d-1}\int d\mathbf{r}\int d\mathbf{v}_1
\int d\mathbf{v}_2\int_{\Omega(\mathbf{r})}d\sig\theta(-\mathbf{v}_{12}\cdot\sig)
\abs{\mathbf{v}_{12}\cdot\sig}
g_2(\mathbf{r}+\boldsymbol{\sigma},\mathbf{r}\lvert n)\nonumber\\
f(\mathbf{r}+\boldsymbol{\sigma},\mathbf{v}_1,t)
f(\mathbf{r},\mathbf{v}_2,t)
\ln\frac{f(\mathbf{r}+\boldsymbol{\sigma},\mathbf{v}_1',t)}
{f(\mathbf{r}+\boldsymbol{\sigma},\mathbf{v}_1,t)},  
\end{eqnarray}
where we have taken into account that 
$g_2(\mathbf{r}+\boldsymbol{\sigma},\mathbf{r}\lvert n)
=g_2(\mathbf{r},\mathbf{r}+\boldsymbol{\sigma}\lvert n)$. Taking into account 
Eq. (\ref{dhk1}) and (\ref{dhk3}), we have 
\begin{eqnarray}
\frac{d\mathcal{H}^{(k)}}{dt}=\frac{\sigma^{d-1}}{2}
\int d\mathbf{r}\int d\mathbf{v}_1
\int d\mathbf{v}_2\int_{\Omega(\mathbf{r})}d\sig\theta(-\mathbf{v}_{12}\cdot\sig)
\abs{\mathbf{v}_{12}\cdot\sig}
g_2(\mathbf{r}+\boldsymbol{\sigma},\mathbf{r}\lvert n)\nonumber\\
f(\mathbf{r}+\boldsymbol{\sigma},\mathbf{v}_1,t)
f(\mathbf{r},\mathbf{v}_2,t)
\ln\frac{f(\mathbf{r}+\boldsymbol{\sigma},\mathbf{v}_1',t)
f(\mathbf{r},\mathbf{v}_2',t)}
{f(\mathbf{r}+\boldsymbol{\sigma},\mathbf{v}_1,t)
f(\mathbf{r},\mathbf{v}_2,t)},  
\end{eqnarray}
that is the expression of the main text. 

Now let us calculate $\frac{d\mathcal{H}^{(c)}}{dt}$. The first contribution is 
\begin{eqnarray}
\frac{d}{dt}\ln \phi(t)=\frac{1}{\phi(t)}\frac{d}{dt}\int d\mathbf{r}_1
w(\mathbf{r}_1,t)\dots\int d\mathbf{r}_Nw(\mathbf{r}_N,t)
\Theta(\mathbf{r}_1,\dots,\mathbf{r}_N)\nonumber\\
=\int d\mathbf{r}\frac{n(\mathbf{r},t)}{w(\mathbf{r},t)}
\frac{\partial}{\partial t}w(\mathbf{r},t), 
\end{eqnarray}
where Eq. (\ref{dw}) has been used. The second contribution is
\begin{eqnarray}
\frac{d}{dt}\int d\mathbf{r}n(\mathbf{r},t)
\ln\frac{n(\mathbf{r},t)}{w(\mathbf{r},t)}
=\int d\mathbf{r}\left[\ln\frac{n(\mathbf{r},t)}{w(\mathbf{r},t)}
\frac{\partial}{\partial t}n(\mathbf{r},t)
-\frac{n(\mathbf{r},t)}{w(\mathbf{r},t)}
\frac{\partial}{\partial t}w(\mathbf{r},t)\right], 
\end{eqnarray}
so that 
\begin{equation}\label{dhc1}
\frac{d\mathcal{H}^{(c)}}{dt}=-\int d\mathbf{r}
\frac{\partial}{\partial t}n(\mathbf{r},t)
\ln\frac{n(\mathbf{r},t)}{w(\mathbf{r},t)}=\int d\mathbf{r}
\ln\frac{n(\mathbf{r},t)}{w(\mathbf{r},t)}\frac{\partial}{\partial\mathbf{r}}
\cdot[n(\mathbf{r},t)\mathbf{u}(\mathbf{r},t)], 
\end{equation}
where the continuity equation, Eq. (\ref{eh1}), has been used. As 
\begin{equation}
\int d\mathbf{r}
\frac{\partial}{\partial\mathbf{r}}\cdot\left[
n(\mathbf{r},t)\mathbf{u}(\mathbf{r},t)
\ln\frac{n(\mathbf{r},t)}{w(\mathbf{r},t)}\right]=
\int_{\partial V} d\mathbf{s}\cdot
\mathbf{u}(\mathbf{r},t)n(\mathbf{r},t)
\ln\frac{n(\mathbf{r},t)}{w(\mathbf{r},t)}=0, 
\end{equation}
because $\mathbf{u}(\mathbf{r},t)\cdot\mathbf{N}(\mathbf{r})=0$ for all 
$\mathbf{r}\in\partial V$, Eq. (\ref{dhc1}) can be written in the form 
\begin{equation}\label{dhc2}
\frac{d\mathcal{H}^{(c)}}{dt}=-\int d\mathbf{r}
n(\mathbf{r},t)\mathbf{u}(\mathbf{r},t)\cdot
\frac{\partial}{\partial\mathbf{r}}\ln\frac{n(\mathbf{r},t)}{w(\mathbf{r},t)}. 
\end{equation}
Now, using the property
\begin{equation}
\frac{\partial}{\partial\mathbf{r}_1}
\theta(\abs{\mathbf{r}_1-\mathbf{r}_2}-\sigma)=
\frac{(\mathbf{r}_1-\mathbf{r}_2)}{\sigma}
\delta(\abs{\mathbf{r}_1-\mathbf{r}_2}-\sigma), 
\end{equation}
we get, from the expressions of $n$ and $n_2$, Eqs. (\ref{dw}) and (\ref{n2w}) 
respectively 
\begin{equation}
\frac{\partial}{\partial\mathbf{r}_1}
\left[\frac{n(\mathbf{r}_1,t)}{w(\mathbf{r}_1,t)}\right]=
\frac{1}{w(\mathbf{r}_1,t)}\int d\mathbf{r}_2
\frac{(\mathbf{r}_1-\mathbf{r}_2)}{\sigma}
\delta(\abs{\mathbf{r}_1-\mathbf{r}_2}-\sigma)
n_2(\mathbf{r}_1,\mathbf{r}_2,t). 
\end{equation}
Performing the pertinent integration to eliminate the delta function, we have
\begin{equation}\label{lnnw}
\frac{\partial}{\partial\mathbf{r}}
\left[\frac{n(\mathbf{r},t)}{w(\mathbf{r},t)}\right]=-\sigma^{d-1}
\frac{n(\mathbf{r},t)}{w(\mathbf{r},t)}\int_{\Omega(\mathbf{r})}d\sig
g_2(\mathbf{r}+\boldsymbol{\sigma},\mathbf{r}\lvert n)
n(\mathbf{r}+\boldsymbol{\sigma},t)\sig . 
\end{equation}
This equation is the generalization of Eq. (\ref{eqne}) for a generic $w$ 
in our non equilibrium ensemble given by Eq. (\ref{rhoN}). By substituting Eq. 
(\ref{lnnw}) into (\ref{dhc2}), we finally obtain 
\begin{eqnarray}
\frac{d\mathcal{H}^{(c)}}{dt}=\sigma^{d-1}\int d\mathbf{r}\int_{\Omega(\mathbf{r})}
d\sig n(\mathbf{r},t) n(\mathbf{r}+\boldsymbol{\sigma},t)\sig\cdot
\mathbf{u}(\mathbf{r},t)g_2(\mathbf{r}+\boldsymbol{\sigma},\mathbf{r}\lvert n)
\nonumber\\
=-\sigma^{d-1}\int d\mathbf{r}\int_{\Omega(\mathbf{r})}
d\sig n(\mathbf{r},t) n(\mathbf{r}+\boldsymbol{\sigma},t)\sig\cdot
\mathbf{u}(\mathbf{r}+\boldsymbol{\sigma},t)
g_2(\mathbf{r}+\boldsymbol{\sigma},\mathbf{r}\lvert n), 
\end{eqnarray}
where, in the last step, we have changed $\sig\to-\sig$ and 
$\mathbf{r}\to\mathbf{r}+\boldsymbol{\sigma}$.

\end{document}